\documentclass[12pt]{article}

\usepackage[utf8]{inputenc}
\usepackage[english]{babel}
\usepackage{caption}
\usepackage{subcaption}
\usepackage{amsmath}
\usepackage{amsfonts}
\usepackage{amssymb}
\usepackage{amsthm}
\usepackage{empheq}
\usepackage{graphicx}
\usepackage{multirow}
\usepackage{float}
\usepackage{verbatim}
\usepackage{psfrag}
\usepackage{geometry}
\usepackage{enumerate}
\usepackage{xcolor}
\usepackage{fancyhdr}
\usepackage[auth-lg]{authblk}
\usepackage[square,numbers]{natbib}
\usepackage{hyperref}
\hypersetup{
    colorlinks=true,
    linkcolor=blue,
    filecolor=magenta,      
    urlcolor=cyan,
    pdfpagemode=FullScreen,
}
\numberwithin{equation}{section}

\newcommand{\be}{\begin{equation}}
\newcommand{\ee}{\end{equation}}
\newcommand{\bea}{\begin{eqnarray}}
\newcommand{\eea}{\end{eqnarray}}
\newcommand{\beano}{\begin{eqnarray*}}
\newcommand{\enano}{\end{eqnarray*}}

\renewcommand{\o}{\overline}

\title{\Large \bf Isomorphism of Analytical Spectrum between Noncommutative Harmonic Oscillator and Landau Problem}
\author[2*]{M. N. Nazmi M. Rusli}
\author[1,3*]{Nurisya M. Shah}
\author[1,3]{Hishamuddin Zainuddin}
\author[1,3]{Chan Kar Tim}
\affil[1]{Laboratory of Computational Sciences and Mathematical Physics, Institute for Mathematical Research (INSPEM), Universiti Putra Malaysia, 43400 UPM Serdang, Selangor, Malaysia}
\affil[2]{Department of Physics, Kulliyyah of Science (KOS), International Islamic University Malaysia (IIUM), 25200 Kuantan, Pahang, Malaysia}
\affil[3]{Department of Physics, Faculty of Science, Universiti Putra Malaysia, 43400 UPM Serdang, Selangor, Malaysia}
\date{\vspace{-1\baselineskip}{\textit{$^*$\small{corresponding email: nazrennazmi98@gmail.com, risya@upm.edu.my}}}}

\fancypagestyle{plain}{%
\fancyhf{} % clear all header and footer fields
\fancyhead[C]{\small{}}
\fancyfoot[C]{\small{\thepage}}

}

\fancypagestyle{fancy2}{
\fancyhf{} % clear all header and footer fields
\lhead{\small}

}

\begin{document}
\renewcommand{\abstractname}{\vspace{-3\baselineskip}}
\pagestyle{plain}
    \begin{@twocolumnfalse}
    \maketitle
	\begin{abstract} \noindent
		The comparison of the Hamiltonians of the noncommutative isotropic harmonic oscillator and Landau problem are analysed to study the specific conditions under which these two models are indistinguishable. The energy eigenvalues and eigenstates of Landau problem in symmetric and two Landau gauges are evaluated analytically. The Hamiltonian of a noncommutative isotropic harmonic oscillator is found by using Bopp's shift in commutative coordinate space. The result shows that the two systems are isomorphic up to the similar values of $n_r$ and $m_l$ and $qB = eB > 0$ for both gauge choices. However, there is an additional requirement for Landau gauge where the noncommutative oscillator has to lose one spatial degree of freedom. It also needs to be parametrized by a factor $\zeta$ for their Hamiltonians to be consistent with each other. The wavefunctions and probability density functions are then plotted and the behaviour that emerges is explained. Finally, the effects of noncommutativity or magnetic field on the eigenstates and their probability distributions of the isomorphic system are shown. \\
		
	\end{abstract}

	\end{@twocolumnfalse}

\thispagestyle{fancy2}

\section{Introduction}

Recently, noncommutative field theories and their derivatives have been a subject of intense research. We refer to \cite{source1,source2,source3} as reading materials to provide some historical background and reviews on the subject over the past few years. Since its establishment, it has invaded various domains of theoretical physics and has effectively continued to evolve and meet the mathematical requirements of various situations. This includes string theory (e.g \cite{source4,source5,source6}), Yang - Mills theory whereby some of its forms is a noncommutative gauge theory (e.g \cite{source7,source8,source9}) and condensed matter theory most notably, quantum Hall effect (e.g \cite{source10}).

A simple insight on the role of noncommutativity in field
theory can be obtained by studying the one particle sector,
which prompted an interest in the study of noncommutative quantum mechanics. The deformation of space due to noncommutativity can be expressed by the commutation relations of Hermitian operators as shall be seen later in Section \ref{sec:3}. Thus, several authors have solved many related problems for example, hydrogen atom\cite{source11}, central potential\cite{source12}, Aharonov - Bohm effect\cite{source13}, Ahoronov - Casher effect\cite{source14}, Klein–Gordon oscillators\cite{source15}, the Landau problem\cite{source16}, and the list goes on.

In \cite{source17}, the relationship between the magnetic field and noncommutativity parameter has been established ($\theta = \frac{4\hbar}{qB}$) for the noncommutative quantum mechanics and the usual Landau problem to be equivalent theories in the lowest Landau levels. A further generalization of this relation was then made in \cite{source18} in the context of relativistic quantum mechanics. In this paper, we would like to argue that the isomorphism between one of the problems of noncommutative quantum mechanics i.e, isotropic harmonic oscillator and Landau problem can also exist in higher-order Landau levels in fact, all Landau levels provided that their specific conditions to be discussed are satisfied. 

The paper is organized as follows. Section \ref{sec:2} is devoted to the energy spectra and wavefunctions of Landau problem in symmetric and Landau gauge by solving the Schr\"{o}dinger equation analytically. Section \ref{sec:3} focuses on finding the solution to the eigenvalue problem of a noncommutative isotropic harmonic oscillator by virtue of comparison of its Hamiltonian and that of Landau problem. Section \ref{sec:4} attempts to explore the behaviour of the wavefunctions and the probability densities plot of the isomorphic system as different quantum numbers are shown. In section \ref{sec:5}, the effects of the magnetic field or noncommutativity are demonstrated on the ground-state eigenfunctions and their probability distributions of the isomorphic system. In the final section, we state our conclusion.

\section{Landau levels} \label{sec:2}
\begin{comment}
The motion of a charged particle in a magnetic field is inherently a problem in a three-dimensional space but it can be remodelled to be in a two-dimensional system since the particle acts as a free particle in the direction of the magnetic field. It is found that the orbital motion is quantized when charged particles are subjected to a magnetic field. Thus, the particles can only occupy orbits with discrete energy eigenvalues called Landau levels~\cite{source20}. We will focus on the orbital motion of the charged particle (hereafter this will be electron) and neglect the spin degree of freedom. Spin will cause a breaking of degeneracy in the presence of a magnetic ﬁeld, owing to the direct coupling with the ﬁeld through the so-called Zeeman term.
\end{comment}

In the presence of a magnetic ﬁeld $\hat{B}$, the canonical momentum $\hat{p}$ is shifted by the magnetic vector potential $\hat{A}$ to give a Hamiltonian of the form
\begin{equation} \label{eq:2.1}
\hat{H} = \frac{\left(\hat{p} - q\hat{A}\right)^2}{2\mu},
\end{equation}
where $q$ and $\mu$ are the charge and the mass of the particle of interest, and $\hat{p} - q\hat{A}$ is the kinetic momentum operator. Hereafter, we adopt $q = -e$ which denotes the charge of an electron. Note that in the presence of a magnetic field, $\hat{p} - q\hat{A}$ represents the true momentum of the particle rather than $\hat{p}$ and is the result of minimal coupling rule \cite{source19}. In this problem, we consider an electron to be freely moving in a two-dimensional system in a uniform magnetic field, $\hat{B} = B\hat{z}$. For such a magnetic field, there are two common gauge choices which will be discussed in greater detail shortly. 

\subsection{Symmetric gauge}
In symmetric gauge, the vector potential can be written as
\begin{equation} \label{eq:2.2}
\hat{A}_x = -\frac{1}{2} B\hat{y}, \quad \hat{A}_y = \frac{1}{2} B\hat{x}.
\end{equation}
Then by substituting \eqref{eq:2.2} into \eqref{eq:2.1},
\begin{equation}
\hat{H} = \frac{1}{2\mu} \left[\left(\hat{p}_x + \frac{qB}{2}\hat{y}\right)^2 + \left(\hat{p}_y - \frac{qB}{2}\hat{x}\right)^2\right].
\end{equation}
Note that we define the z-axis such that $qB > 0$. For the case of the electron where $q < 0$, it means that the magnetic field is along the negative z-axis. Alternatively, $qB > 0$ is also true for the proton where  $q > 0$ when the magnetic field is along the positive z-axis. Hence we will now adopt $qB = eB$ instead, where $eB$ is also greater than $0$. After a few algebraic steps,
\begin{equation}  \label{eq:2.4}
\hat{H} = \frac{1}{2\mu}\left(\hat{p}_x^2 + \hat{p}_y^2\right) + \frac{e^2B^2}{8\mu}\left(\hat{x}^2 + \hat{y}^2\right) - \frac{eB}{2\mu}\left(\hat{x}\hat{p}_y - \hat{y}\hat{p}_x\right),
\end{equation}
where
\begin{equation}
\hat{L}_z \equiv \hat{x} \hat{p}_y - \hat{y} \hat{p}_x,
\end{equation}
is the z-component of the orbital angular momentum operator. Let
\begin{equation}
\omega_c = \frac{eB}{\mu},
\end{equation}
which represents the cyclotron frequency in SI units.

However in some works in literature such as~\cite{source21,source24}, the Hamiltonian can also be written as 
\begin{equation}
\hat{H} = \frac{1}{2\mu} \left[\left(\hat{p}_x + \frac{qB}{2c}\hat{y}\right)^2 + \left(\hat{p}_y - \frac{qB}{2c}\hat{x}\right)^2\right],
\end{equation}
which leads to 
\begin{equation}
\hat{H} = \frac{1}{2\mu}\left(\hat{p}_x^2 + \hat{p}_y^2\right) + \frac{e^2B^2}{8\mu c^2}\left(\hat{x}^2 +\hat{y}^2\right) - \frac{eB}{2\mu c}\left(\hat{x}\hat{p}_y - \hat{y}\hat{p}_x\right).
\end{equation} 
As a result,
\begin{equation}
\omega_c = \frac{eB}{\mu c},
\end{equation}
which still represents the cyclotron frequency but in \textit{Gaussian} units as the Lorentz force differs by a factor of $1/c$ where $c$ is the speed of light. Both are valid representations of the Hamiltonian for the system. In this work, the Hamiltonian which considers the SI units of the cyclotron frequency will be considered. 

Returning to our initial problem of calculating the eigenstates and eigenvalues of \eqref{eq:2.4},
\begin{comment}
we first note that since $[\hat{H},\hat{L}_z] = 0$, we can write the eigenstates $\Psi(r,\varphi)$ as
\begin{align} \label{eq:2.10}
\Psi(r,\varphi) & = R(r)\Phi(\varphi) \nonumber
\\
& = R(r) e^{i m_l \varphi)}
\end{align}
where $m_l$ is the eigenvalue of the $\hat{L}_z$ operator. Using periodic boundary conditions in the variable $\varphi$, $\Psi(r,\varphi + 2\pi) = \Psi(r,\varphi)$ requires $m_l$ to be an integer, and the eigenvalues of $\hat{L}_z$ are thus quantized. The resulting  radial ordinary differential equation is
\begin{equation} \label{eq:2.11}
\left[-\frac{\hbar^2}{2\mu} \left(\frac{d^2}{d r^2} + \frac{1}{r}\frac{d}{d r}\right) + \frac{m_l^2 \hbar^2}{2\mu r^2} + \frac{e^2B^2}{8\mu} r^2 - \frac{m_l\hbar eB}{2\mu}\right]R(r) = ER(r).
\end{equation}
\eqref{eq:2.11} is precisely of the form of a two - dimensional  isotropic radial quantum harmonic oscillator Hamiltonian with
frequency $\frac{\omega_c}{2}$, with an energy shift of $-\frac{m_l \hbar eB}{2\mu}$.
\end{comment}
by applying a few change of variables and orthogonality relation, the solution to the time-independent Schr\'odinger equation is 
\begin{align} \label{eq:2.10}
\Psi_{n_r,m_l}(r,\varphi) = & \frac{1}{\sqrt{2\pi}} \left(\frac{eB}{\hbar}\right)^{\frac{1}{2}} \sqrt{\frac{n_r!}{(n_r + |m_l|)!}} \left(\frac{eB}{2\hbar}r^2\right)^\frac{|m_l|}{2} \nonumber
\\
& \exp \left(-\frac{eB}{4\hbar}r^2\right) L_{n_r}^{|m_l|} \left(\frac{eB}{2\hbar}r^2\right) e^{im_l\varphi}.
\end{align}
where $n_r$ and $m_l$ are radial and angular momentum quantum numbers respectively \cite{source25} and is in agreement with \cite{source26}. As for the eigenvalues, they are given as
\begin{align}
E_{n_r,m_l} & = \left(2n_r + |m_l| + 1\right)\hbar\frac{\omega_c}{2} - m_l\hbar\frac{\omega_c}{2},\\
& = \left(2n_r + |m_l| + 1\right)\hbar\frac{eB}{2\mu} - m_l\hbar\frac{eB}{2\mu}.
\end{align}
The first term mimics the energy eigenvalues of the standard two-dimensional harmonic oscillator (with the frequency $\frac{\omega_c}{2}$ instead). The second term comes from the coupled momentum-gauge term that is manifested in the Hamiltonian after expanding the kinetic momentum operator. This term represents the interaction between the canonical momenta and the magnetic field, and what distinguishes this problem from a standard planar isotropic oscillator~\cite{source26}.

Now, the interesting question that could be pondered is, what if we choose to use $qB = -eB$ instead? At first, this may look trivial but a careful calculation reveals that the resulting energy levels are actually different from the case that we obtain previously. In fact, it can be shown that
\begin{align}
E_{n_r,m_l} & = \left(2n_r + |m_l| + 1\right)\hbar\frac{\omega_c}{2} + m_l\hbar\frac{\omega_c}{2},\\
& = \left(2n_r + |m_l| + 1\right)\hbar\frac{eB}{2\mu} + m_l\hbar\frac{eB}{2\mu},
\end{align}
while the eigenfunction maintains the same form as in \eqref{eq:2.10}. Notice that the sign of the second term is different than before which implies that, the mere assumption that we make on the product, $qB$ can actually change the pattern of degeneracy. We refer to \cite{source26} for a detailed derivation of the eigenstates and eigenenergies obtained when $qB = -eB$.

\begin{comment}
\begin{table}[H]
	\centering
	\caption{Ordered pair of quantum numbers and its corresponding energy for $m_l \leq 0$}
		\begin{tabular}{|p{5cm}|p{7cm}|}
			\hline
			Energy (in units of $\hbar\frac{\omega_c}{2}$)& $(n_r,m_l)$\\
			\hline
			$7$ & $(3,0),(3,-1),(3,-2)...$ \\ 
			$5$ & $(2,0)(2,-1)(2,-2),...$ \\ 
			$3$ & $(1,0)(1,-1)(1,-2),...$ \\ 
			$1$ & $(0,0),(0,-1),(0,-2),...$ \\ 
			\hline
		\end{tabular} 
	\label{tab:3}
\end{table}

\begin{table}[H]
	\centering
	\caption{Ordered pair of quantum numbers and its corresponding energy for $m_l > 0$}
		\begin{tabular}{|p{5cm}|p{7cm}|}
			\hline
			Energy (in units of $\hbar\frac{\omega_c}{2}$)& $(n_r,m_l)$\\
			\hline
			$7$ & $(0,3)(1,2)(2,1)$ \\ 
			$5$ & $(0,2)(1,1)$ \\ 
			$3$ & $(0,1)$ \\ 
			\hline
		\end{tabular} 
	\label{tab:4}
\end{table}
\end{comment}

\begin{comment}
The argument is in varying degree similar to what has been presented except that we have to take into account on how the different signs of $m_l$ affects the system. The energy eigenstates are exactly similar to the one derived previously.
\end{comment}

\subsection{Landau gauge}
As for the Landau gauge, it can be written in two different ways;
\begin{equation} \label{eq:2.17}
\mbox{First Landau gauge : } \hat{A}_x = - B\hat{y},	\hat{A}_y = 0,
\end{equation}
\begin{equation} \label{eq:2.18}
\mbox{Second Landau gauge : } \hat{A}_x = 0,	\hat{A}_y = B\hat{x}.
\end{equation}
From here thereon, we are only going to focus our attention on the first Landau gauge, as the discussion that follows will also be applied to the second Landau gauge in a similar manner to the first Landau gauge but with a minute difference as shall be pointed out later. Then by substituting \eqref{eq:2.17} into \eqref{eq:2.1}
\begin{equation}
\hat{H} = \frac{1}{2\mu} \left[\left(\hat{p}_x + qB\hat{y}\right)^2 + \left(\hat{p}_y\right)^2\right],
\end{equation}

\begin{comment}
\begin{equation}
\mbox{Second Landau gauge : }\hat{H} = \frac{1}{2\mu} \left[\left(\hat{p}_x\right)^2 + \left(\hat{p}_y - qB\hat{x}\right)^2\right].
\end{equation}
\end{comment}
Just like in the symmetric gauge, we will define the z-axis such that $qB > 0$ and hence, $qB = eB$. After a few algebraic steps,
\begin{equation} \label{eq:2.20}
\hat{H} = \frac{1}{2\mu}\left(\hat{p}_x^2 + \hat{p}_y^2\right) + \frac{e^2B^2}{2\mu}\hat{y}^2 + \frac{eB}{\mu}\hat{y}\hat{p}_x,
\end{equation}
\begin{comment}
Immediately, we see that $[\hat{H},\hat{p}_x] = 0$. Thus, we can separate our eigenstate as $\Psi(x,y) \propto X(x)Y(y)$ to get
\begin{equation} \label{eq:2.21}
\Psi(x,y) \propto e^{ik_x x} Y(y), 
\end{equation}
where the operator $\hat{p}_x$ is replaced by its eigenvalue $\hbar k_x$. Substituting \eqref{eq:2.20} into the stationary Schr\"{o}dinger equation and the fact that the equation will be separable based on \eqref{eq:2.21}, we will have 
\begin{equation} \label{eq:2.22}
\left[-\frac{\hbar^2}{2\mu}\frac{d^2}{dy^2} + \frac{1}{2}\mu\omega_c\left(y + \frac{\hbar k_x}{eB}\right)^2\right]Y(y) = EY(y),
\end{equation}
\end{comment}
The derivation of the solution of the Schr\"{o}dinger equation involving the Hamiltonian in \eqref{eq:2.20} is very well known as it has the mathematical structure of a shifted harmonic oscillator~\cite{source27} and thus, will not be repeated here. It is
\begin{align}
\Psi_{n_y,k_x}(x,y) = & \frac{1}{\sqrt{2^{n_y} n_y!}}\left(\frac{eB}{\pi\hbar}\right)^{\frac{1}{4}}\exp\left[-\frac{eB}{2\hbar}\left(y + \frac{\hbar k_x}{eB}\right)^2\right] \nonumber
\\
& H_n\left[\frac{eB}{\hbar}\left(y + \frac{\hbar k_x}{eB}\right)\right] e^{ik_x x},
\end{align}
\begin{comment}
\begin{align}
\mbox{Second Landau gauge : } \Psi_{n_x,k_y}(x,y) = & \frac{1}{\sqrt{2^{n_x} n_x!}}\left(\frac{eB}{\pi\hbar}\right)^{\frac{1}{4}}\exp\left[-\frac{eB}{2\hbar}\left(x - \frac{\hbar k_y}{eB}\right)^2\right] \nonumber
\\
& H_n\left[\frac{eB}{\hbar}\left(x - \frac{\hbar k_y}{eB}\right)\right] e^{ik_y y},
\end{align}
\end{comment}
where $H_n$ is the Hermite polynomial. This method of solution has introduced two quantum numbers; $k_x$, a real number, and $n_y$, a non-negative integer. This latter quantum number counts the number of nodes of the probability density. On the other hand, the eigenvalues are
\begin{equation} \label{eq:2.24}
E_{n_y,k_x} = \left(2n_y + 1\right)\hbar \frac{\omega_c}{2},
\end{equation}
\begin{comment}
\begin{equation}
\mbox{Second Landau gauge : } E_{n_x,k_y} = \left(2n_x + 1\right)\hbar \frac{\omega_c}{2}.
\end{equation}
\end{comment}
Note the lack of dependence on the quantum number $k_x$. Consequently, the energies are infinitely degenerate for every non-negative integer $n_y$. The energy eigenvalues of the second Landau gauge will also take the form of \eqref{eq:2.24}. In regard to the energy eigenstates, we need to change the sign of the shifted term of the position of the harmonic oscillator to have $-\frac{\hbar k_y}{eB}$ with quantum numbers, $k_y$ and $n_x$. In contrast to our result, in \cite{source26}, an assumption of $qB = -eB$ is analysed instead. The only difference of our results with those in \cite{source26} is, we need to change the sign of the spatial shift of the eigenfunction of the oscillator in both Landau gauges while having the same energy. 

\begin{comment}
Nevertheless, we will still mention them here as follows; for the energy eigenstates
\begin{align}
\mbox{First Landau gauge : } \Psi_{n_y,k_x}(x,y) = & \frac{1}{\sqrt{2^{n_y} n_y!}}\left(\frac{eB}{\pi\hbar}\right)^{\frac{1}{4}}\exp\left[-\frac{eB}{2\hbar}\left(y - \frac{\hbar k_x}{eB}\right)^2\right] \nonumber
\\
& H_n\left[\frac{eB}{\hbar}\left(y - \frac{\hbar k_x}{eB}\right)\right] e^{ik_x x},
\end{align}
\begin{align}
\mbox{Second Landau gauge : } \Psi_{n_x,k_y}(x,y) = & \frac{1}{\sqrt{2^{n_x} n_x!}}\left(\frac{eB}{\pi\hbar}\right)^{\frac{1}{4}}\exp\left[-\frac{eB}{2\hbar}\left(x + \frac{\hbar k_y}{eB}\right)^2\right] \nonumber
\\
& H_n\left[\frac{eB}{\hbar}\left(x + \frac{\hbar k_y}{eB}\right)\right] e^{ik_y y}.
\end{align}
As for the energy eigenvalues they are just identical to the case of when $qB > 0$ and hence, we will not write it here.
\end{comment}

\section{Noncommutative isotropic harmonic oscillator} \label{sec:3}
The Hamiltonian of a particle of mass $m$ which oscillates with an angular frequency $\omega$ under the influence of an isotropic harmonic oscillator potential in the noncommutative space is formulated as
\begin{equation} \label{eq:3.1}
\hat{H} = \frac{1}{2m} \left(\mathbf{\hat{p}}_1^2 + \mathbf{\hat{p}}_2^2\right) + \frac{1}{2} m\omega^2 \left(\mathbf{\hat{x}}_1^2 + \mathbf{\hat{x}}_2^2\right),
\end{equation}

There are various formulations of quantum mechanics on noncommutative Moyal phase spaces that can be used to deal with related problems such as the above. Among them include canonical, path-integral, Weyl-Wigner, and systematic formulations (see \cite{source28}). We will utilize canonical formulation in our work. Then, by applying Bopp's shift transformation, the non-commuting coordinates can be expressed in terms of commuting coordinates in the following form 
\begin{equation} \nonumber
\mathbf{\hat{x}}_i = \hat{x}_i - \frac{\Theta}{2\hbar}\hat{p}_j,
\end{equation} 
\begin{equation} \label{eq:3.2}
\mathbf{\hat{p}}_i = \hat{p}_i,
\end{equation}
We can see from the above that $\mathbf{\hat{p}}_i$ maps onto itself.
In the sequel, we often take $\Theta_{ij} = \theta\varepsilon_{ij}$ where $\Theta$ is the constant, frame-dependent parameters and $\varepsilon_{ij}$ is the Levi-Civita symbol~\cite{source29} and they are connected as written as follows
\begin{equation} \label{eq:3.3}
	\varepsilon_{ij}=
    \begin{cases}
    	1 & i < j\\
        0 & i = j\\
        -1 & i > j.
    \end{cases}
\end{equation}
The commutators involving position and momentum in \eqref{eq:3.2} are $[\mathbf{\hat{x}}_i,\mathbf{\hat{x}}_j] = i\Theta_{ij}$, $[\mathbf{\hat{p}}_i,\mathbf{\hat{p}}_j] = 0$ and $[\hat{x}_i,\hat{p}_j] = i\hbar\delta_{ij}$. Due to the transformation, the new variables will satisfy the usual commutation relations below
\begin{align}
\left[\hat{x}_i,\hat{x}_j\right] &= 0,\nonumber\\
\left[\hat{x}_i,\hat{p}_j\right] &= i\hbar\delta_{ij},\\
\left[\hat{p}_i,\hat{p}_j\right] &= 0.\nonumber
\end{align}

With the new variables defined in \eqref{eq:3.2}, the Hamiltonian in the ordinary space will then be
\begin{equation}
\hat{H} = \frac{1}{2m} \left[\hat{p}_1^2 + \hat{p}_2^2\right] + \frac{1}{2} m\omega^2 \left[\left(\hat{x}_1 - \frac{\theta}{2\hbar} \hat{p}_2 \right)^2 + \left(\hat{x}_2 + \frac{\theta}{2\hbar} \hat{p}_1 \right)^2\right].
\end{equation}
After a few algebraic steps,
\begin{equation}
\hat{H} = \left(\frac{1}{2m} + \frac{m\omega^2\theta^2}{8\hbar^2}\right) \left(\hat{p}_1^2 + \hat{p}_2^2\right) + \frac{1}{2} m\omega^2 \left(\hat{x}_1^2 + \hat{x}_2^2\right) - \frac{\theta}{2\hbar} m\omega^2 \left(\hat{x}_1 \hat{p_2} - \hat{x}_2 \hat{p}_1\right).
\end{equation}
Let $\hat{x}_1 = \hat{x}$ and $\hat{x}_2 = \hat{y}$ then,
\begin{equation} \label{eq:3.7}
\hat{H} = \left(\frac{1}{2m} + \frac{m\omega^2\theta^2}{8\hbar^2}\right) \left(\hat{p}_x^2 + \hat{p}_y^2\right) + \frac{1}{2} m\omega^2 \left(\hat{x}^2 + \hat{y}^2\right) - \frac{\theta}{2\hbar} m\omega^2 \left(\hat{x} \hat{p}_y - \hat{y} \hat{p}_x\right),
\end{equation}
where
\begin{equation}
\hat{L}_z \equiv \hat{x}_1 \hat{p}_2 - \hat{x}_2 \hat{p}_1,
\end{equation}
is the z-component of the orbital angular momentum operator. It is obvios that the third term will vanish once we set $\theta = 0$ and is analogue to that corresponding to the Landau problem on  $\mathbb{R}^2$~\cite{source24}. Now, let
\begin{equation}
\gamma = \frac{\theta}{2\hbar}M\Omega^2.
\end{equation} 
Then, rearrange \eqref{eq:3.7} by introducing the effective mass and frequency so that the first two terms have the form of the Hamiltonian of an isotropic harmonic oscillator as follows
\begin{equation} \label{eq:3.10}
\hat{H} = \frac{1}{2M} \left(\hat{p}_x^2 + \hat{p}_y^2\right) + \frac{1}{2} M\Omega^2 \left(\hat{x}^2 + \hat{y}^2\right) - \gamma \hat{L}_z,
\end{equation}
where
\begin{equation}
\frac{1}{2M} \equiv \frac{1}{2m} + \frac{m\omega^2\theta^2}{8\hbar^2},\nonumber
\end{equation}  
\begin{equation} 
M = \frac{m}{1 + \frac{m^2\omega^2\theta^2}{4\hbar^2}}.
\end{equation}
As a result
\begin{equation}
M\Omega^2 \equiv m\omega^2, \nonumber
\end{equation}
\begin{equation}
\Omega^2 = m\omega^2 \left(\frac{1 + \frac{m^2\omega^2\theta^2}{4\hbar^2}}{m}\right), \nonumber
\end{equation}
\begin{equation}
\Omega = \omega \sqrt{1 + \frac{m^2\omega^2\theta^2}{4\hbar^2}}.
\end{equation}
where the definition of the parameters above are also used in \cite{source22,source23}. Here comes the main question in this paper, can we somehow find the energy eigenvalues and eigenstates of the noncommutative oscillator from the mathematical framework of Landau problem? This shall be explored in the following subsections.

\subsection{Symmetric gauge}
From Section \ref{sec:2}, we have determined the energy spectra and wavefunctions of Landau levels. The question now is, what are the conditions to be obeyed in order for Landau levels and noncommutative oscillator to be equivalent? We can naively compare the Hamiltonian of the two systems, \eqref{eq:3.10} and \eqref{eq:2.4} then try to infer from there. Thus we have,
\begin{equation}
\frac{m}{1 + \frac{m^2\omega^2\theta^2}{4\hbar^2}} = \mu, \nonumber
\end{equation}
\begin{equation}
\frac{1}{2} m\omega^2 = \frac{e^2 B^2}{8\mu}, \nonumber\\
\end{equation}
\begin{equation} \label{eq:3.13}
\frac{\theta}{2\hbar} m\omega^2 = \frac{eB}{2\mu}.
\end{equation}
Realize that, the set of equations that we have from the comparison of the coefficients of the Hamiltonians needs to be true simultaneously. In other words, \eqref{eq:3.13} altogether has to be consistent but, that is not the case here. For the two systems to be isomorphic, we need to invoke some mathematical scheme by redefining $M$ and $\Omega$ in other terms instead of $m$ and $\omega$. This is allowable as $M$ and $\Omega$ are respectively the parameters that characterize the effective mass and frequency of the oscillator in a commutative space. With reference to \cite{source17} which is generalized for any noncommutative quantum mechanical problem in central field, we are going to let the effective mass and frequency to be
\begin{equation} \label{eq:3.14}
M = \frac{2\hbar^2}{\theta^2},
\end{equation} 
and
\begin{equation} \label{eq:3.15}
\Omega = \frac{\theta}{\hbar},
\end{equation}
which are $\theta$-dependent. Due to \eqref{eq:3.14} and \eqref{eq:3.15}, the resulting Hamiltonian of the noncommutative oscillator is the following,
\begin{equation} \label{eq:3.16}
\hat{H} = \frac{\theta^2}{4\hbar^2} \left(\hat{p_x}^2 + \hat{p_y}^2\right) + \left(\hat{x}^2 + \hat{y}^2\right) - \frac{\theta}{\hbar} \hat{L_z}.
\end{equation}
Realize that by imposing the oscillator to have the new effective mass and frequency in terms of $\theta$ alone, the Hamiltonian \eqref{eq:3.16} lack of parameters permits an opportunity of introducing a factor to make the two systems consistent. By multiplying each of the coefficients of the Hamiltonian by a factor $\zeta$ to be determined, then
\begin{equation} 
\hat{H} = \frac{\zeta\theta^2}{4\hbar^2} \left(\hat{p_x}^2 + \hat{p_y}^2\right) + \zeta\left(\hat{x}^2 + \hat{y}^2\right) - \frac{\zeta\theta}{\hbar} \hat{L_z}.
\end{equation}
Now, by comparing the newly-defined Hamiltonian of the noncommutative oscillator and the Landau levels, hence
\begin{equation}
\frac{\zeta\theta^2}{4\hbar^2} = \frac{1}{2\mu}, \nonumber
\end{equation}
\begin{equation}    	
\zeta = \frac{e^2 B^2}{8\mu}, \nonumber
\end{equation}
\begin{equation} \label{eq:3.18}
\frac{\zeta\theta}{\hbar} = \frac{eB}{2\mu}.
\end{equation}
Unlike the comparison from \eqref{eq:3.13}, a close inspection reveals that \eqref{eq:3.18} is altogether consistent and in accordance with~\cite{source17}. As a result, together with factor $\zeta$, the effective mass and frequency of the oscillator will then be equal to those of Landau levels as shown below
\begin{align} \label{eq:3.19}
\frac{1}{2M} & = \frac{\zeta\theta^2}{4\hbar^2}, \nonumber\\
M & = \frac{2\hbar^2}{\zeta\theta^2},
\end{align} 
\begin{align} \label{eq:3.20}
\frac{1}{2}M\Omega^2 & = \zeta, \nonumber\\
\Omega & = \frac{\zeta\theta}{\hbar},
\end{align}
where $\zeta = \frac{e^2 B^2}{8\mu}$ is a controlling factor. By calculation, we can evaluate that when Landau levels have $\mu = m_e = 9.109 \times 10^{-31}$ and $\omega_c = \frac{eB}{\mu} = 2.110 \times 10^{12}$, then the oscillator would have  $M = 4.620 \times 10^{-37}$ and $\Omega = 2.081 \times 10^{18}$ due to \eqref{eq:3.14} and \eqref{eq:3.15}. Since we have established the relationship above, we can always examine a different mass and frequency of either of the systems and be able to find the same parameters of the other system studied for them to be isomorphic.
 
Let us contemplate the effect of a constant $B$ and $\theta$ on their respective systems. In the quantum Hall effect experiments, the magnetic fields are typically about $B = 12 T$ ~\cite{source17,source30, source31}. Using the second and third equations in \eqref{eq:3.18},
\begin{equation} \label{eq:3.21}
\theta = \frac{4\hbar}{eB} =  2.195 \times 10^{-16},
\end{equation}
in $m^2$ or it can also be written as $\theta = 0.2195 \times 10^{-11}$ in $cm^2$ which is in conformance with \cite{source17}. For this value of $\theta$, one cannot distinguish between noncommutative quantum mechanics (in this case, it would be the noncommutative isotropic harmonic oscillator) and the usual Landau problem. 

Since we have determined the value of $\theta$, we should also compute the value of the controlling factor $\zeta$, that is
\begin{equation} \label{eq:3.22}
\zeta = \frac{e^2 B^2}{8\mu} = 5.071 \times 10^{-7},
\end{equation}
which implies that just like $\theta$, there is a dependence on the charge of the particle and the magnetic field which acts on it in the Landau problem with an additional dependence on the mass of the particle.

Hence, since their Hamiltonians are now similar, analysing Landau levels is tantamount to analysing the noncommutative oscillator. Let us revisit the energy eigenvalues of Landau levels in symmetric gauge for $qB = eB > 0$. They are
\begin{equation} \label{eq:3.23}
E_{n_r,m_l} = \left(2n_r + |m_l| + 1\right)\hbar\frac{eB}{2\mu} - m_l\hbar\frac{eB}{2\mu}.
\end{equation}
For the noncommutative oscillator, we can express \eqref{eq:3.23} in terms of $\theta$ and $\zeta$ as follows
\begin{equation} \label{eq:3.24}
E_{n_r,m_l} = \left(2n_r + |m_l| + 1\right)\zeta\theta - m_l\zeta\theta.
\end{equation}

Therefore, we conclude that the noncommutative oscillator and Landau levels are isomorphic provided that the pair of quantum numbers $n_r$ and $m_l$ of both systems have the same values. Besides, the Hamiltonian of the noncommutative oscillator has to be parametrized by the controlling factor $\zeta$ so that it is consistent with the Hamiltonian of Landau levels in symmetric gauge. In regards to the wavefunctions of Landau levels, we recall that
\begin{align} \label{eq:3.25}
\Psi_{n_r,m_l}(r,\varphi) = & \frac{1}{\sqrt{2\pi}} \left(\frac{eB}{\hbar}\right)^{\frac{1}{2}} \sqrt{\frac{n_r!}{(n_r + |m_l|)!}} \left(\frac{eB}{2\hbar}r^2\right)^\frac{|m_l|}{2} \nonumber
\\
& \exp \left(-\frac{eB}{4\hbar}r^2\right) L_{n_r}^{|m_l|} \left(\frac{eB}{2\hbar}r^2\right) e^{im_l\varphi}.
\end{align}
For the noncommutative oscillator, again by expressing \eqref{eq:3.25} in terms of $\theta$ and $\zeta$, one obtains
\begin{align} \label{eq:3.26}
\Psi_{n_r,m_l}(r,\varphi) = & \frac{1}{\sqrt{2 \pi}} \left(\frac{4}{\theta}\right)^\frac{1}{2} \sqrt{\frac{n_r!}{(n_r + |m_l|)!}} \left(\frac{2}{\theta}r^2\right)^\frac{|m_l|}{2} \nonumber
\\
& \exp \left(-\frac{1}{\theta}r^2\right) L_{n_r}^{|m_l|} \left(\frac{2}{\theta}r^2\right) e^{im_l\varphi}.
\end{align}
With the above results, we can find the analytical spectra of the noncommutative oscillator in the context of Landau problem in a symmetric gauge. However, the results obtained are by no means representing the general spectra of the solutions of the noncommutative oscillator but rather specific to the case when $M$ and $\Omega$ are $\theta-$dependent designated in \eqref{eq:3.19} and \eqref{eq:3.20}. We refer to \cite{source32,source33,source34,source35} as reading materials that deliver more elaborate research on the spectra of the  noncommutative oscillator using path integral formulation.

The same question as raised before can still be asked here. What if we impose the following condition, $qB = -eB$? We will encounter a situation where the comparison cannot be made as
\begin{equation} \label{eq:3.27}
-\frac{\theta}{2\hbar} = \frac{eB}{2\mu},
\end{equation}
which is clearly false as the left-hand side is negative while the right-hand side is positive. This is the result when comparing the third term of their Hamiltonians. One might argue that why do we not account for this with a new $\zeta$. The problem is, the $\zeta-$ term is also present in the first and second terms of the Hamiltonian of the noncommutative oscillator. If we change them, it will generate a false statement like in \eqref{eq:3.27}. Hence, we can say that this is also a condition for isomorphism. We want to emphasize again that the results of the energy spectra and wavefunctions of the noncommutative oscillator that we obtain here are applied only when $M$ and $\Omega$ are $\theta-$dependent which is the limitation of this formalism. 

\subsection{Landau gauge}
As for the two Landau gauges, we cannot directly compare the coefficients of their Hamiltonians as their variables or operators are clearly different from one another. Nonetheless, an assumption can be made for each to account for this. For the first Landau gauge, assume that the position operator, $\hat{x}$ to be the zero operator, $\hat{0}$. The same supposition goes to the second Landau gauge as well but with $\hat{y}$. Hence, the correlation can then be made as follows,
\begin{equation}
\frac{m}{1 + \frac{m^2\omega^2\theta^2}{4\hbar^2}} = \mu, \nonumber
\end{equation}
\begin{equation}
\frac{1}{2} m\omega^2 = \frac{e^2 B^2}{2\mu}, \nonumber\\
\end{equation}
\begin{equation} \label{eq:4.17}
\frac{\theta}{2\hbar} m\omega^2 = \frac{eB}{\mu}.
\end{equation}

Notice that the set of equations in \eqref{eq:4.17} is reminiscent to the one in symmetric gauge and therefore, we will not repeat the discussion here. However we will show how their Hamiltonians take the form as portrayed below,
\begin{table}[H]
	\centering
	\caption{Comparison between the Hamiltonian of Landau gauges and noncommutative oscillator}
	\large
		\begin{tabular}{|p{6cm}|p{8cm}|}
			\hline
			{\normalsize Hamiltonian of Landau gauge} & {\normalsize Hamiltonian of noncommutative oscillator}\\
			\hline
			$\frac{1}{2\mu}\left(\hat{p}_x^2 + \hat{p}_y^2\right) + \frac{e^2B^2}{2\mu}\hat{y}^2 + \frac{eB}{\mu}\hat{y}\hat{p}_x$ & $\frac{\zeta\theta^2}{4\hbar^2} \left(\hat{p_x}^2 + \hat{p_y}^2\right) + \zeta\left(\hat{x}^2 + \hat{y}^2\right) - \frac{\zeta\theta}{\hbar} \hat{L_z}$ \\ 
			$\frac{1}{2\mu}\left(\hat{p}_x^2 + \hat{p}_y^2\right) + \frac{e^2B^2}{2\mu}\hat{x}^2 - \frac{eB}{\mu}\hat{x}\hat{p}_y$ & $\frac{\zeta\theta^2}{4\hbar^2} \left(\hat{p_x}^2 + \hat{p_y}^2\right) + \zeta\left(\hat{x}^2 + \hat{y}^2\right) - \frac{\zeta\theta}{\hbar} \hat{L_z}$ \\ 
			\hline
		\end{tabular} 
	\label{tab:5}
\end{table}
After solving the time-independent Schr\"{o}dinger equation, the energy eigenvalues of the noncommutative oscillator in the context of the first Landau gauge are 
\begin{equation}
E_{n_y} = \left(2n_y + 1\right)\zeta\theta
\end{equation}
and the energy eigenstate is
\begin{align} \label{eq:3.29}
\Psi_{n_y}^{(k_0)}(y) = & \frac{1}{\sqrt{2^{n_y} n_y!}}\left(\frac{4}{\pi\theta}\right)^{\frac{1}{4}}\exp\left[-\frac{2}{\theta}\left(y + \frac{\theta}{4} k_0\right)^2\right] \nonumber
\\
& H_n\left[\frac{4}{\theta}\left(y + \frac{\theta}{4} k_0\right)\right] 
\end{align}
\begin{comment}
\begin{equation}
\mbox{Second Landau gauge : } E_{n_x} = \left(2n_x + 1\right)\zeta\theta
\end{equation}
\end{comment}
The important thing to note here is, the system is more restrictive compared to the case of symmetric gauge since we have to lose a spatial degree of freedom. As a result, the system is nondegenerate for every quantum number, $n_y$. The real number $k_x$ will be a constant that we will denote as $k_0$, since the system has lost the degree of freedom associated with position $x$. This is equivalent to the statement that the position has localized to a fixed value. Consequently, it implies that it is impossible to determine $k_0$ as it is correlated with the momentum $p_x$ of the particle due to the uncertainty principle. We will treat $k_0$ as a parameter instead of a quantum number and its effect on the system will be analysed in Section \ref{sec:4.2}. If we use the second Landau gauge, the energy eigenvalues resemble the ones obtained before and, the energy eigenfunction will have a negative shift in position. Otherwise, as before, it will not work if we assume $qB < 0$.

\section{Effect of quantum numbers on the isomorphic system} \label{sec:4}

Since the conditions under which these theories are isomorphic have been laid out in the previous section, we will plot the wavefunctions and probability density functions for the first few lower quantum numbers to observe their effects on the isomorphic system. Let us substitute $\theta = 2.195 \times 10^{-16}$ that is defined from the previous section into \eqref{eq:3.26} for symmetric gauge and \eqref{eq:3.29} for the first Landau gauge. Since their eigenstates are identical (under certain assumptions) as proven in the previous section, plotting the wavefunction of a noncommutative oscillator is tantamount to that corresponding to Landau levels. Hence, plotting a noncommutative oscillator alone is sufficient to avoid redundancy in the upcoming figures and discussion. 

\subsection{Symmetric gauge}

By using \textit{Mathematica}, the 3D plot of one of the ground-state wavefunctions and its probability density of the isomorphic system when both quantum numbers are zero are shown as follows,

\begin{figure}[H]
	\centering
		\includegraphics[scale=0.5]{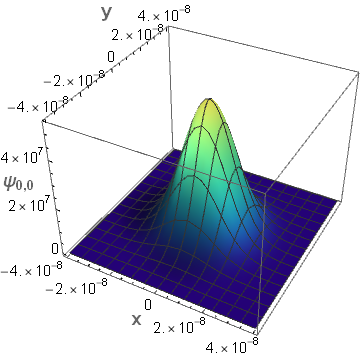}
		\caption{Ground-state, $\Psi_{0,0}(r,\varphi)$ of the isomorphic system with respect to symmetric gauge} 
		\label{fig:1}
\end{figure}

\begin{figure}[H]
	\centering
		\includegraphics[scale=0.5]{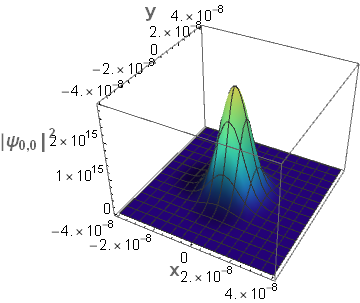}
		\caption{Ground-state, $|\Psi_{0,0}(r,\varphi)|^2$ of the isomorphic system with respect to symmetric gauge} 
		\label{fig:2}
\end{figure}

In Figure \ref{fig:1}, since $n_r$ and $m_l$ are both zero, the imaginary part of the wavefunction naturally vanishes and so are other related terms in \eqref{eq:3.26} leaving the wavefunction to have a structure of a $e^{-r^2}$ function. The same argument goes to Figure \ref{fig:2} as well except that it follows the structure of a $e^{-2r^2}$ which explains its slighly different feature. Yet, the characteristic shape for both follows the two-dimensional Gaussian function. Due to its distribution, it is most likely to find the particle at the origin. Generally, the wavefunction bears no physical interpretation while the probability density function does as its own name implies,  by giving us the likelihood of finding an electron (or some other system) at some given point in space based on the Born rule. Hence, in the upcoming discussion, the wavefunction (in this paper we will focus on its real part) will only serve as a visualisation while the probability distribution is subjected to further analysis. The ground-state energy is $1.112 \times 10^{-22} J$ or $6.943 \times 10^{-4} eV$.

\subsubsection{Effect of $n_r$ on $\Psi_{n_r,0}$} \label{sec:4.1.1}
The exact magnitude of energy for an increasing value of the radial quantum number $n_r$ can be seen as follows. 
\begin{table}[H]
	\centering
	\caption{Ordered pair of quantum numbers and its corresponding energy of the isomorphic system in the units of $J$ and $eV$}
		\begin{tabular}{|p{2cm}|p{5cm}|p{5cm}|}
			\hline
			$(n_r,m_l)$ & Energy (in $J$) & Energy (in $eV$) \\
			\hline
			$(1,0)$ & $3.340 \times 10^{-22}$ & $2.085 \times 10^{-3}$ \\ 
			$(2,0)$ & $5.566 \times 10^{-22}$ & $3.475 \times 10^{-3}$ \\ 
			$(3,0)$ & $7.793 \times 10^{-22}$ & $4.864 \times 10^{-3}$ \\ 
			\hline
		\end{tabular} 
	\label{tab:6}
\end{table}

\begin{figure}[H]
\centering
\begin{subfigure}{.5\textwidth}
  \centering
	  \includegraphics[width=0.8\linewidth]{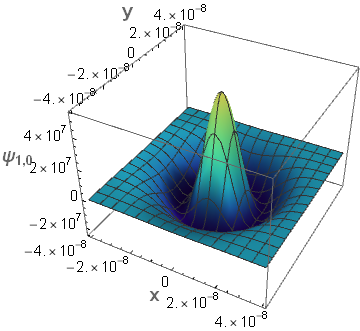}
	  \caption{First excited state, $\Psi_{1,0}(r,\varphi)$ }
\end{subfigure}\hfill
\begin{subfigure}{.5\textwidth}
  \centering
	  \includegraphics[width=0.8\linewidth]{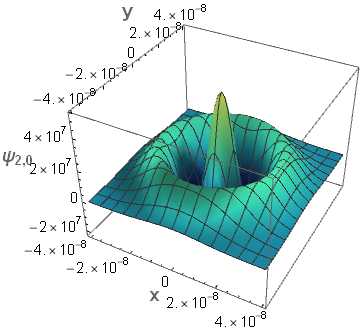}
	  \caption{Second excited state, $\Psi_{2,0}(r,\varphi)$}
\end{subfigure}\hfill
\begin{subfigure}{.5\textwidth}
  \centering
	  \includegraphics[width=0.8\linewidth]{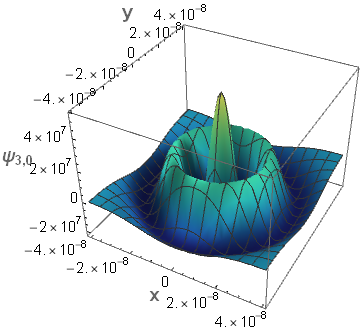}
	  \caption{Third excited state, $\Psi_{3,0}(r,\varphi)$ }
\end{subfigure}\hfill
\caption{Effect of $n_r$ on $\Psi_{n_r,0}(r,\varphi)$ of the isomorphic system with respect to symmetric gauge}
\label{fig:3}
\end{figure}

\begin{figure}[H]
\centering
\begin{subfigure}{.5\textwidth}
  \centering
	  \includegraphics[width=0.8\linewidth]{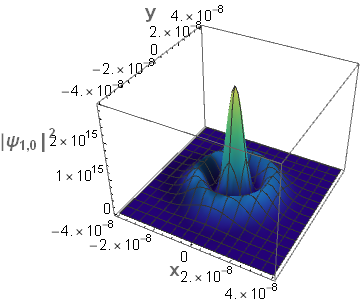}
	  \caption{First excited state, $|\Psi_{1,0}(r,\varphi)|^2$ }
\end{subfigure}\hfill
\begin{subfigure}{.5\textwidth}
  \centering
	  \includegraphics[width=0.8\linewidth]{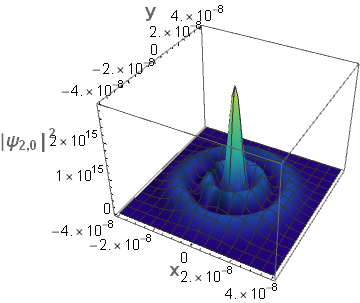}
	  \caption{Second excited state, $|\Psi_{2,0}(r,\varphi)|^2$}
\end{subfigure}\hfill
\begin{subfigure}{.5\textwidth}
  \centering
	  \includegraphics[width=0.8\linewidth]{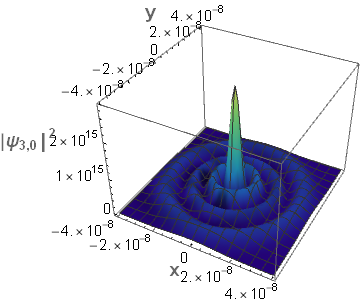}
	  \caption{Third excited state, $|\Psi_{3,0}(r,\varphi)|^2$ }
\end{subfigure}\hfill
\caption{Effect of $n_r$ on $|\Psi_{n_r,0}(r,\varphi)|^2$ of the isomorphic system with respect to symmetric gauge}
\label{fig:4}
\end{figure}

Figure \ref{fig:3} shows that in a fixed domain, the number of concentric  rings formed that can be seen on the nodes and antinodes of the wavefunctions increases as $n_r$ increases. This is actually reasonable because when we observe Figure \ref{fig:4}, those antinodes will turn to nodes and it is in accordance with what $n_r$ actually represents. It is a positive integer that corresponds to the number of nodes in the wave function moving radially out from the center. Just like $|\Psi_{0,0}|^2$, the likelihood of finding the particle at the origin is also the highest. Though as $m_l$ increases, so does the number of nodes causing the probability distribution to be smeared out to the existing nodes. The more the number of nodes, the lower their resulting amplitudes.

\subsubsection{Effect of  $m_l$ on $Re[\Psi_{0,m_l}]$} \label{sec:4.1.2}
For an increasing value of the angular momentum quantum number $m_l$,  the exact value of the energy levels will be evaluated by using the parameters mentioned before.
\begin{table}[H]
	\centering
	\caption{Ordered pair of quantum numbers and its corresponding energy of the isomorphic system in the units of $J$ and $eV$}
		\begin{tabular}{|p{2cm}|p{5cm}|p{5cm}|}
			\hline
			$(n_r,m_l)$ & Energy (in $J$) & Energy (in $eV$) \\
			\hline
			$(0,3)$ & $1.112 \times 10^{-22}$ & $6.943 \times 10^{-4}$ \\ 
			$(0,2)$ & $1.112 \times 10^{-22}$ & $6.943 \times 10^{-4}$ \\ 
			$(0,1)$ & $1.112 \times 10^{-22}$ & $6.943 \times 10^{-4}$ \\ 
			$(0,-1)$ & $3.340 \times 10^{-22}$ & $2.085 \times 10^{-3}$ \\ 
			$(0,-2)$ & $5.566 \times 10^{-22}$ & $3.475 \times 10^{-3}$ \\ 
			$(0,-3)$ & $7.793 \times 10^{-22}$ & $4.864 \times 10^{-3}$ \\ 
			\hline
		\end{tabular} 
	\label{tab:7}
\end{table}

\begin{figure}[H]
\centering
\begin{subfigure}{.45\textwidth}
  \centering
	  \includegraphics[width=0.8\linewidth]{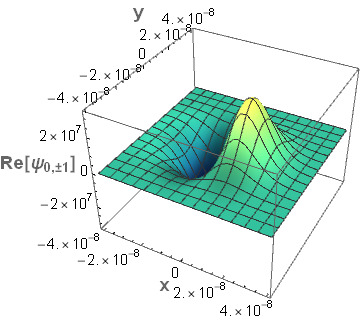}
	  \caption{Ground-state, $Re[\Psi_{0,1}(r,\varphi)]$ and first excited state, $Re[\Psi_{0,-1}(r,\varphi)]$}
\end{subfigure}\hfill
\begin{subfigure}{.45\textwidth}
  \centering
	  \includegraphics[width=0.8\linewidth]{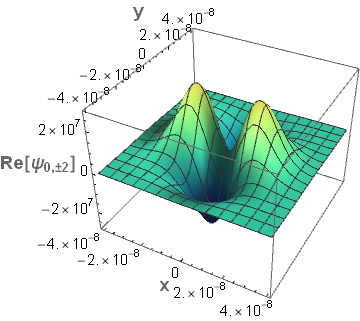}
	  \caption{Ground-state, $Re[\Psi_{0,2}(r,\varphi)]$ and second excited state, $Re[\Psi_{0,-2}(r,\varphi)]$}
\end{subfigure}\hfill
\begin{subfigure}{.45\textwidth}
  \centering
	  \includegraphics[width=0.8\linewidth]{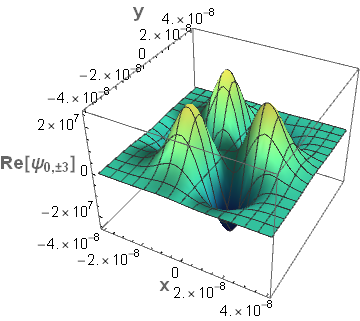}
	  \caption{Ground-state, $Re[\Psi_{0,3}(r,\varphi)]$ and third excited state, $Re[\Psi_{0,-3}(r,\varphi)]$}
\end{subfigure}\hfill
\caption{Effect of $m_l$ on $Re[\Psi_{0,m_l}(r,\varphi)]$ on the isomorphic system with respect to symmetric gauge}
\label{fig:5}
\end{figure}

\begin{figure}[H]
\centering
\begin{subfigure}{.45\textwidth}
  \centering
	  \includegraphics[width=0.8\linewidth]{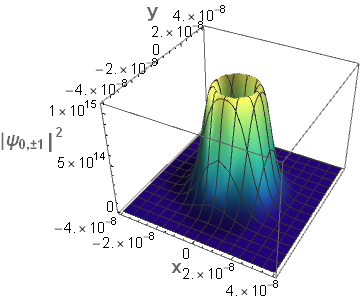}
	  \caption{Ground-state, $|\Psi_{0,1}(r,\varphi)|^2$ and first excited state, $|\Psi_{0,-1}(r,\varphi)|^2$}
\end{subfigure}\hfill
\begin{subfigure}{.45\textwidth}
  \centering
	  \includegraphics[width=0.8\linewidth]{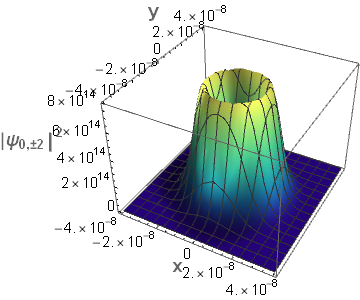}
	  \caption{Ground-state, $|\Psi_{0,2}(r,\varphi)|^2$ and second excited state, $|\Psi_{0,-2}(r,\varphi)|^2$}
\end{subfigure}\hfill
\begin{subfigure}{.45\textwidth}
  \centering
	  \includegraphics[width=0.8\linewidth]{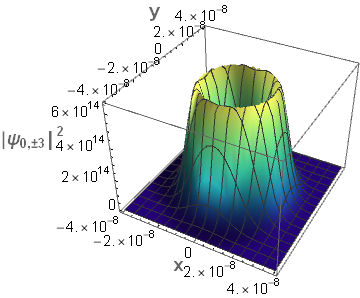}
	  \caption{Ground-state, $|\Psi_{0,3}(r,\varphi)|^2$ and third excited state,  $|\Psi_{0,-3}(r,\varphi)|^2$}
\end{subfigure}\hfill
\caption{Effect of $m_l$ on $|\Psi_{0,m_l}(r,\varphi)|^2$ on the isomorphic system with respect to symmetric gauge}
\label{fig:6}
\end{figure}

Notice that unlike Figures \ref{fig:1} until \ref{fig:4}, there are two orders of state to which each of the plots in Figure \ref{fig:5} and \ref{fig:6} belongs. This comes from the rotational symmetry associated with states with opposite signs of $m_l$ and thus, producing identical geometrical shapes. When $n_r = 0$, all states with $m_l \geq 0$ stay at the ground level. The remaining states with $m_l < 0$ will mimic the behaviour of $n_r$ in Section \ref{sec:4.1.1} as documented in Table \ref{tab:7}. Figure \ref{fig:5} displays the real part of the wavefunction and we can see that the number of local minima and maxima designated  by the troughs and valleys formed increases as $|m_l|$ increases. In addition, Figure \ref{fig:6} indicates that just like Figure \ref{fig:4}, the distribution is radially symmetric and the probability of finding a particle farther away from the origin is greater as $|m_l|$ increases since the radius at which the maximum amplitude is located increases. Unlike in the previous subsection, we only plot the real part of the wavefunctions but not as a whole. This is because, when $m_l \neq 0$, we cannot plot the real and imaginary part of the wavefunctions on the same set of axes.

\subsubsection{Effect of both $n_r$ and $m_l$ on $Re[\Psi_{n_r,m_l}]$}
When both $n_r$ and $m_l$ are nonzero, for the sake of demonstration, Table \ref{tab:8} depict the energy levels of both systems.
\begin{table}[H]
	\centering
	\caption{ Ordered pair of quantum numbers and its corresponding energy of the noncommutative oscillator in the units of $J$ and $eV$}
		\begin{tabular}{|p{2cm}|p{5cm}|p{5cm}|}
			\hline
			$(n_r,m_l)$ & Energy (in $J$) & Energy (in $eV$) \\
			\hline
			$(1,1)$ & $3.340 \times 10^{-22}$ & $2.085 \times 10^{-3}$ \\ 
			$(2,1)$ & $5.566 \times 10^{-22}$ & $3.475 \times 10^{-3}$ \\ 
			$(3,1)$ & $7.793 \times 10^{-22}$ & $4.864 \times 10^{-3}$ \\ 
			$(1,2)$ & $3.340 \times 10^{-22}$ & $2.085 \times 10^{-3}$ \\ 
			$(2,2)$ & $5.566 \times 10^{-22}$ & $3.475 \times 10^{-3}$ \\ 
			$(3,2)$ & $7.793 \times 10^{-22}$ & $4.864 \times 10^{-3}$ \\ 
			$(1,3)$ & $3.340 \times 10^{-22}$ & $2.085 \times 10^{-3}$ \\ 
			$(2,3)$ & $5.566 \times 10^{-22}$ & $3.475 \times 10^{-3}$ \\ 
			$(3,3)$ & $7.793 \times 10^{-22}$ & $4.864 \times 10^{-3}$ \\  
			$(1,-1)$ & $5.566 \times 10^{-22}$ & $3.475 \times 10^{-3}$ \\ 
			$(2,-1)$ & $7.793 \times 10^{-22}$ & $4.864 \times 10^{-3}$ \\ 
			$(3,-1)$ & $1.002 \times 10^{-21}$ & $6.254 \times 10^{-3}$ \\ 
			$(1,-2)$ & $7.793 \times 10^{-22}$ & $4.864 \times 10^{-3}$ \\ 
			$(2,-2)$ & $1.002 \times 10^{-21}$ & $6.254 \times 10^{-3}$ \\ 
			$(3,-2)$ & $1.225 \times 10^{-21}$ & $7.644 \times 10^{-3}$ \\ 
			$(1,-3)$ & $1.002 \times 10^{-21}$ & $6.254 \times 10^{-3}$ \\ 
			$(2,-3)$ & $1.225 \times 10^{-21}$ & $7.644 \times 10^{-3}$ \\ 
			$(3,-3)$ & $1.447 \times 10^{-21}$ & $9.034 \times 10^{-3}$ \\
			\hline
		\end{tabular}
	\label{tab:8}
\end{table}

\begin{figure}[H]
\centering
\begin{subfigure}{.5\textwidth}
  \centering
	  \includegraphics[width=0.8\linewidth]{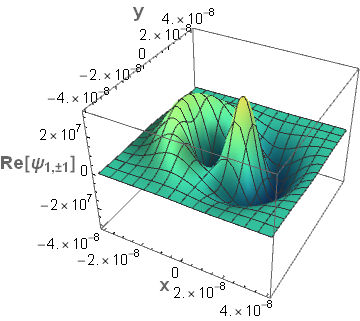}
	  \caption{$Re[\Psi_{1,\pm 1}(r,\varphi)]$}
\end{subfigure}\hfill
\begin{subfigure}{.5\textwidth}
  \centering
	  \includegraphics[width=0.8\linewidth]{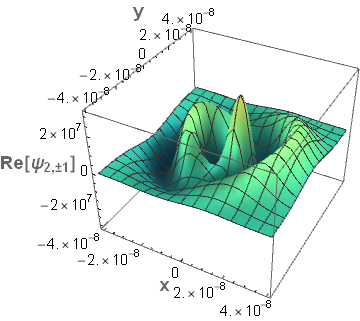}
	  \caption{$Re[\Psi_{2,\pm 1}(r,\varphi)]$}
\end{subfigure}\hfill
\begin{subfigure}{.5\textwidth}
  \centering
	  \includegraphics[width=0.8\linewidth]{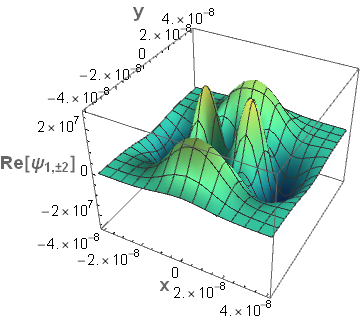}
	  \caption{$Re[\Psi_{1,\pm 2}(r,\varphi)]$}
\end{subfigure}\hfill
\begin{subfigure}{.5\textwidth}
  \centering
	  \includegraphics[width=0.8\linewidth]{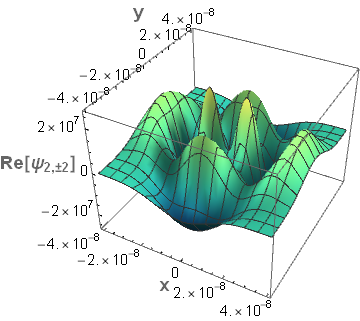}
	  \caption{$Re[\Psi_{2,\pm 2}(r,\varphi)]$}
\end{subfigure}\hfill
\caption{Effect of both $n_r$ and $m_l$ on $Re[\Psi_{n_r,m_l}(r,\varphi)]$ on the isomorphic system with respect to symmetric gauge}
\label{fig:7}
\end{figure}

\begin{figure}[H]
\centering
\begin{subfigure}{.5\textwidth}
  \centering
	  \includegraphics[width=0.8\linewidth]{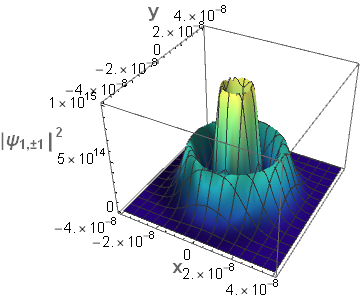}
	  \caption{$|\Psi_{1,\pm 1}(r,\varphi)|^2$}
	  \label{fig:8a}
\end{subfigure}\hfill
\begin{subfigure}{.5\textwidth}
  \centering
	  \includegraphics[width=0.8\linewidth]{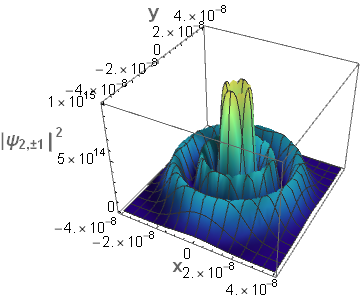}
	  \caption{$|\Psi_{2,\pm 1}(r,\varphi)|^2$}
	  \label{fig:8b}
\end{subfigure}\hfill
\begin{subfigure}{.5\textwidth}
  \centering
	  \includegraphics[width=0.8\linewidth]{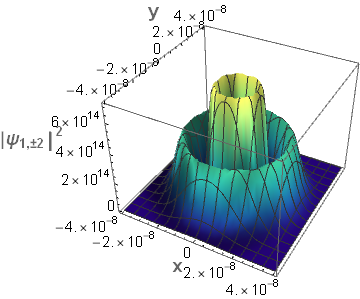}
	  \caption{$|\Psi_{1,\pm 2}(r,\varphi)|^2$}
	  \label{fig:8c}
\end{subfigure}\hfill
\begin{subfigure}{.5\textwidth}
  \centering
	  \includegraphics[width=0.8\linewidth]{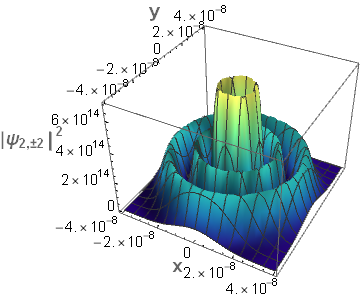}
	  \caption{$|\Psi_{2,\pm 2}(r,\varphi)|^2$}
	  \label{fig:8d}
\end{subfigure}\hfill
\caption{Effect of both $n_r$ and $m_l$ on $|\Psi_{n_r,m_l}(r,\varphi)|^2$ on the isomorphic system with respect to symmetric gauge}
\label{fig:8}
\end{figure}

Figure \ref{fig:7} shows the complex features of some of the arbitrarily chosen states with specific values of $n_r$ and $m_l$ respectively to illustrate its combined effect (due to them being nonzero) on the real part of the wavefunctions. Let us divert our attention to the probability distribution plot. Realize that the distribution of Figure \ref{fig:8c} seems to be spread out farther than the origin while maintaining the same structure compared to Figure \ref{fig:8a}. This is not surprising as Figure \ref{fig:8c} has a higher $|m_l|$ by an addend of 1. As for when comparing Figure \ref{fig:8a} and \ref{fig:8b}, the latter has an additional node which is also anticipated as it has a higher value of $n_r$ by an addend of 1. By combining these two features, we will have Figure \ref{fig:8d}. This behaviour can also be extended to the higher-order state of the isomorphic system. We do not specify the order of state for each plot as the ideas to determine it, are straightforward and be extended from Section \ref{sec:4.1.1} and Section \ref{sec:4.1.2}.

\subsection{Landau gauge} \label{sec:4.2}
For this subsection, we study the effect of the quantum number $n_y$ and a parameter $k_0$ on the wavefunctions and probability distributions of the isomorphic system in the context of the first Landau gauge. Even though the functions are two-dimensional in nature as they only depend on the position $y$, we decide to let our $k_0$ to be the variable of our functions to study their collective effects (with $n_y$) on the same set of axes. This will not create any issue on the interpretation of the plot as the effect of $k_0$ is trivial as shall be seen later. Recall that $k_0$ is also unpredictable and hence, we will not provide a table to show some exact value of energies.

%By using \textit{Mathematica}, 
The contour plot of the aforementioned functions are provided in Figure \ref{fig:9} and \ref{fig:10} below,

\begin{figure}[H]
\centering
\begin{subfigure}{.5\textwidth}
  \centering
	  \includegraphics[width=0.8\linewidth]{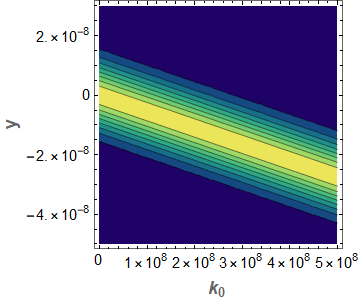}
	  \caption{$\Psi_0^{(k_0)}(y)$}
\end{subfigure}\hfill
\begin{subfigure}{.5\textwidth}
  \centering
	  \includegraphics[width=0.8\linewidth]{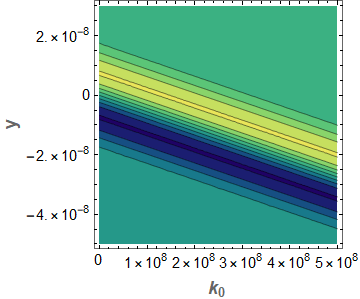}
	  \caption{$\Psi_1^{(k_0)}(y)$}
\end{subfigure}\hfill
\begin{subfigure}{.5\textwidth}
  \centering
	  \includegraphics[width=0.8\linewidth]{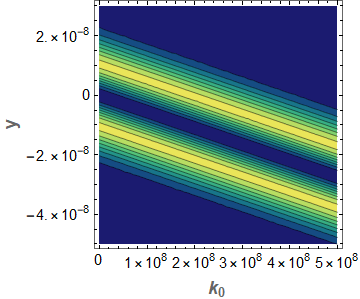}
	  \caption{$\Psi_2^{(k_0)}(y)$}
\end{subfigure}\hfill
\caption{Effect of $n$ on $\Psi_{n_y}^{(k_0)}(y)$ of the isomorphic system with respect to the first Landau gauge and varying parameter $k_0$}
\label{fig:9}
\end{figure}

\begin{figure}[H]
\centering
\begin{subfigure}{.5\textwidth}
  \centering
	  \includegraphics[width=0.8\linewidth]{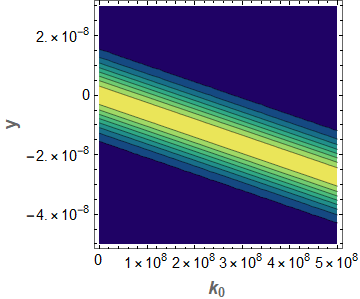}
	  \caption{$|\Psi_0^{(k_0)}(y)|^2$}
\end{subfigure}\hfill
\begin{subfigure}{.5\textwidth}
  \centering
	  \includegraphics[width=0.8\linewidth]{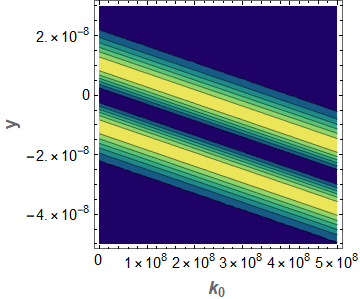}
	  \caption{$|\Psi_1^{(k_0)}(y)|^2$}
\end{subfigure}\hfill
\begin{subfigure}{.5\textwidth}
  \centering
	  \includegraphics[width=0.8\linewidth]{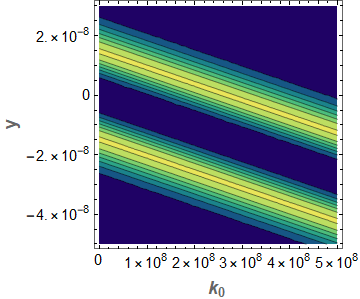}
	  \caption{$|\Psi_2^{(k_0)}(y)|^2$}
\end{subfigure}\hfill
\caption{Effect of $n$ on $|\Psi_{n_y}^{(k_0)}(y)|^2$ of the isomorphic system with respect to the  first Landau gauge and varying parameter $k_0$}
\label{fig:10}
\end{figure}

From Figures \ref{fig:9} and \ref{fig:10}, we can see that the parameter $k_0$ only plays a role in shifting the harmonic oscillator to the right of the origin as $k_0$ increases. From the mathematical form of the wavefunction, we can think of the system as a shifted harmonic oscillator where the behaviour of such function is well understood. By right, the quantum number $n_y$ reflects the number of nodes and antinodes of the wavefunction and only the number of nodes for the probability density curve. However, we can clearly see that only two nodes are present even when $n_y$ is greater than 1 as shown in \ref{fig:10}. This is due to the factor used in the wavefunction which makes the nodes look invisible. Besides, the amplitude of the two highest peaks designated by the brightest strips on the plot also increases as $n_y$ increases which implies that the amplitude of the other seemingly invisible peaks in between should decrease and thus, making it even harder to observe. This behaviour is also displayed in higher-order states.

\section{Effect of $B$ or $\theta$ on the isomorphic system}\label{sec:5}

In this section, the effect of the magnetic field or noncommutativity on the isomorphic system will be presented. We will also use the parameters that have been defined in the previous section for plotting purposes. 
\begin{comment}
\begin{table}[H]
	\centering
		\begin{tabular}{|p{2cm}|p{4cm}|p{4cm}|}
			\hline
			$B$ (in $T$) & $\theta$ (in $m^2$) & $\zeta$\\
			\hline
			$10$ & $2.634 \times 10^{-16}$ & $3.522 \times 10^{-7}$\\ 
			\hline
			$15$ & $1.756 \times 10^{-16}$ & $7.924 \times 10^{-7}$\\ 
			\hline
			$20$ & $1.317 \times 10^{-16}$ & $1.409 \times 10^{-6}$\\ 
			\hline
		\end{tabular} 
	\caption{Magnetic field strength, noncommutativity parameter and controlling factor}
	\label{tab:10}
\end{table}
\end{comment} 

\begin{comment}
From Table \ref{tab:10}, we find out that even a minute change in $\theta$ could dramatically change $B$ to show that $B$ is very sensitive to change due to $\theta$. Though practically speaking it is more appropriate to say the other way around that is, a  higher increase in $B$ leads to a noticeable effect of the decrease in $\theta$ since after all, we can only control $B$ and not $\theta$. $\zeta$ is in the units of $C^2 \cdot T^2 \cdot kg^{-1}$ or $kg \cdot Hz^2$. These two are dimensionally equivalent as the cyclotron frequency can also be in the units of $C \cdot T \cdot kg^{-1}$.
\end{comment} 

\subsection{Symmetric gauge} 
 
\begin{figure}[H]
\centering
\begin{subfigure}{.5\textwidth}
  \centering
	  \includegraphics[width=0.8\linewidth]{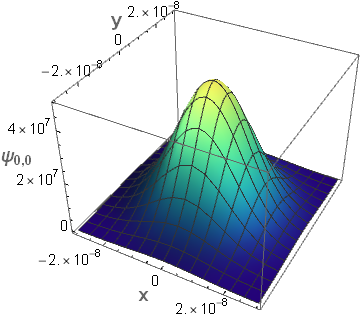}
	  \caption{$\Psi_{0,0}(r,\varphi)$ at $B = 10$T}
\end{subfigure}\hfill
\begin{subfigure}{.5\textwidth}
  \centering
	  \includegraphics[width=0.8\linewidth]{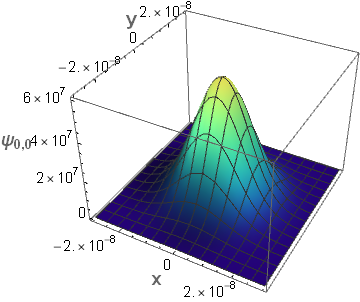}
	  \caption{$\Psi_{0,0}(r,\varphi)$ at $B = 15$T}
\end{subfigure}\hfill
\begin{subfigure}{.5\textwidth}
  \centering
	  \includegraphics[width=0.8\linewidth]{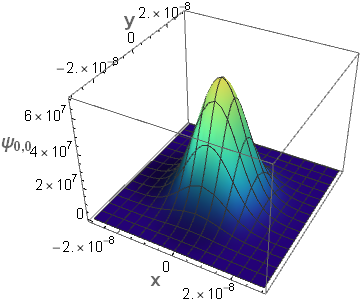}
	  \caption{$\Psi_{0,0}(r,\varphi)$ at $B = 20$T}
\end{subfigure}\hfill
\caption{Effect of $B$ or $\theta$ on the function $\Psi_{0,0}(r,\varphi)$ with respect to symmetric}
\label{fig:11}
\end{figure}

\begin{figure}[H]
\centering
\begin{subfigure}{.5\textwidth}
  \centering
	  \includegraphics[width=0.8\linewidth]{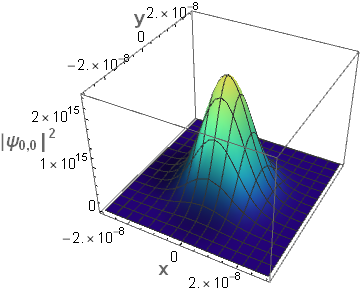}
	  \caption{$\Psi_{0,0}(r,\varphi)$ at $B = 10$T}
\end{subfigure}\hfill
\begin{subfigure}{.5\textwidth}
  \centering
	  \includegraphics[width=0.8\linewidth]{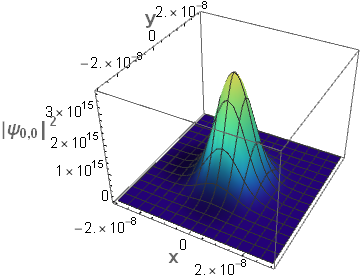}
	  \caption{$\Psi_{0,0}(r,\varphi)$ at $B = 15$T}
\end{subfigure}\hfill
\begin{subfigure}{.5\textwidth}
  \centering
	  \includegraphics[width=0.8\linewidth]{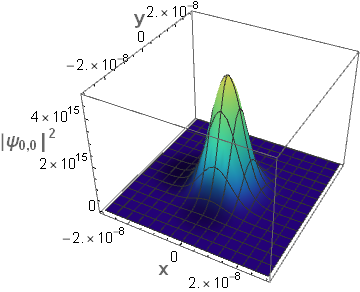}
	  \caption{$\Psi_{0,0}(r,\varphi)$ at $B = 20$T}
\end{subfigure}\hfill
\caption{Effect of $B$ or $\theta$ on probability density, $|\Psi_{0,0}(r,\varphi)|^2$ with respect to symmetric}
\label{fig:12}
\end{figure}

Let us consider the following random values of the magnetic field; $10T$, $15T$ and $20T$. From Figures \ref{fig:11} and \ref{fig:12}, we only plot the ground-state wavefunctions and their respective probability densities i.e, $\Psi_{0,0}$ and $|\Psi_{0,0}|^2$ as the effect can naturally be extended to higher-order states. We observe that the radius at which the distribution begins to flatten out decreases as $B$ increases or $\theta$ decreases. This is due to the radial symmetry of the wavefunction. To compensate for the reduction in radius, the amplitude has to increase to maintain the norm to be one.
 
\subsection{Landau gauge}

This time we also assign the same values of the magnetic field ($10T$, $15T$ and $15T$). We only plot the ground-state wavefunctions and probability distributions just like before. Figures \ref{fig:13} and \ref{fig:14} below show that the deviation of the harmonic oscillator due to parameter $k_0$ from the origin to be lesser as the magnetic field increases or noncommutativity decreases. This is expected as these two parameters are present in the shifted position term in \eqref{eq:3.29}. The same behaviour is also present in higher-order states.

\begin{figure}[H]
\centering
\begin{subfigure}{.5\textwidth}
  \centering
	  \includegraphics[width=0.8\linewidth]{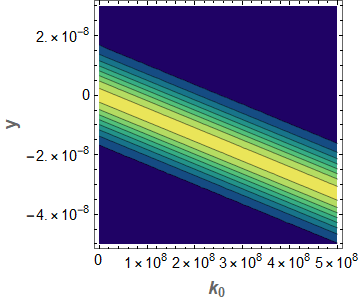}
	  \caption{$\Psi_0^{(k_0)}(y)$ at $B = 5T$}
\end{subfigure}\hfill
\begin{subfigure}{.5\textwidth}
  \centering
	  \includegraphics[width=0.8\linewidth]{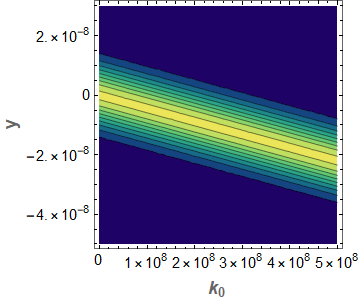}
	  \caption{$\Psi_1^{(k_0)}(y)$ at $B = 10T$}
\end{subfigure}\hfill
\begin{subfigure}{.5\textwidth}
  \centering
	  \includegraphics[width=0.8\linewidth]{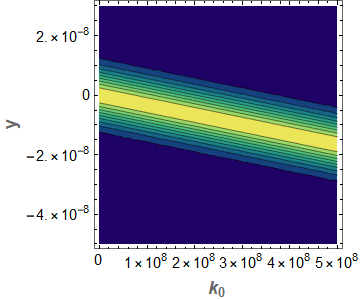}
	  \caption{$\Psi_2^{(k_0)}(y)$ at $B = 15T$}
\end{subfigure}\hfill
\caption{Effect of $B$ or $\theta$ on the function $\Psi_{n_y}^{(k_0)}(y)$ of the isomorphic system with respect to the first Landau gauge and varying parameter $k_0$}
\label{fig:13}
\end{figure}

\begin{figure}[H]
\centering
\begin{subfigure}{.5\textwidth}
  \centering
	  \includegraphics[width=0.8\linewidth]{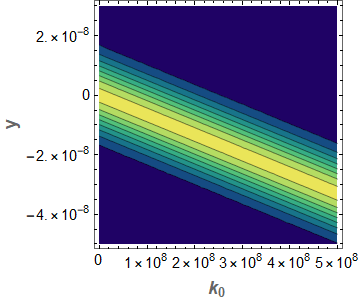}
	  \caption{$|\Psi_0^{(k_0)}(y)|^2$ at $B = 5T$}
\end{subfigure}\hfill
\begin{subfigure}{.5\textwidth}
  \centering
	  \includegraphics[width=0.8\linewidth]{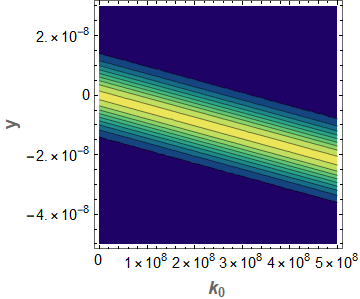}
	  \caption{$|\Psi_1^{(k_0)}(y)|^2$ at $B = 10T$}
\end{subfigure}\hfill
\begin{subfigure}{.5\textwidth}
  \centering
	  \includegraphics[width=0.8\linewidth]{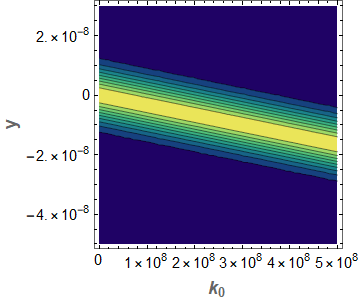}
	  \caption{$|\Psi_2^{(k_0)}(y)|^2$ at $B = 15T$}
\end{subfigure}\hfill
\caption{Effect of $B$ or $\theta$ on the probability density, $|\Psi_{n_y}^{(k_0)}(y)|^2$ of the isomorphic system with respect to the  first Landau gauge and varying parameter $k_0$}
\label{fig:14}
\end{figure}

\section{Conclusions}

In this work, we propose a mathematical formalism as an approach to solve noncommutative harmonic oscillator problems within the framework of the Landau problem. We argue that the equivalence of noncommutative quantum mechanics in the central field and the Landau problem as portrayed by \cite{source12} for the lowest Landau level can be extended to all Landau levels in the case of a harmonic oscillator. We find that the noncommutative isotropic harmonic oscillator and Landau levels can be considered as an isomorphic system in the context of both Landau and symmetric gauges by satisfying a number of conditions. In the context of symmetric gauge, they are isomorphic up to the similar values of $n_r$ and $m_l$ for $qB = eB > 0$. The noncommutative oscillator also has to be parametrized by a factor $\zeta$ to make the Hamiltonians consistent with each other. As for Landau gauge, the requirements are the same but with an addition that, the noncommutative oscillator has to lose one spatial degree of freedom. 

With this research, the problems involving the analytical spectrum of the noncommutative harmonic oscillator can now be understood within the framework of the Landau problem. This framework provides a simpler and more generalized approach to the noncommutative oscillator problems since the Landau problems are very well understood and it also applies to all energy levels. It is also of major interest to study if this relation still holds when a gauge-invariant Landau problem with some potential is considered. One can also imagine the possibility of isomorphism if the Hamiltonian is energy-dependent. 

\section*{Acknowledgments}

The authors wish to acknowledge the financial support provided through the IPM-UPM Grant, Project No. 9645700 Universiti Putra Malaysia.

\end{document}